\documentclass{PoS}

\usepackage{subcaption} % side by side figures etc
\usepackage{multirow} % multirow/column tables
\usepackage[separate-uncertainty = true]{siunitx} % units, angles etc. - does not work with deluxetable ... :(
\newcommand{\similar}{\ensuremath{\sim}}

\newcommand{\GRs}{$\gamma$-ray }

\newcommand*\diff{\mathop{}\!\mathrm{d}}

\title{Raster Scanning the Crab Nebula to Produce an Extended VHE Calibration Source}

\ShortTitle{Raster Scanning the Crab Nebula to Produce an Extended VHE Calibration Source}

\author{\speaker{Ralph Bird} for the VERITAS Collaboration\thanks{http://veritas.sao.arizona.edu}\\
        University College Dublin\\
        E-mail: \email{ralph.bird.1@gmail.com}}

\abstract{The Crab Nebula has long been the standard reference point source for very-high-energy (VHE, E $>\SI{100}{GeV}$) gamma-ray observatories such as VERITAS.  
It has enabled testing and improvement of analysis methods, validation of techniques, and has served as a calibration source.  
No comparable extended source is known with a high, constant flux and well understood morphology.  
In order to artificially generate such a source, VERITAS has performed raster scans across the Crab Nebula.  
By displacing the source within the field-of-view in a known pattern, it is possible to generate an extended calibration source for verification of extended source analysis techniques.  
The method as well as early results of this novel technique are presented.}

\FullConference{The 34th International Cosmic Ray Conference,\\
		30 July- 6 August, 2015\\
		The Hague, The Netherlands}

\begin{document}

\section{Introduction}
Since its initial discovery as a very-high-energy (VHE, $>$\SI{100}{GeV}) \GRs source, the Crab Nebula has proved to be a very useful calibration source for the verification of systems and processes and the development of new techniques.
Its main advantages are that it is bright, with a constant flux (at least to a level that variability has thus far not been detected by any imaging atmospheric Cherenkov Telescope (IACT) \cite{OFDB2015}) and visible by all the major observatories (allowing for verification between them).
Since the majority of sources observed by IACTs are point sources, the Crab Nebula (which appears to IACTs as a point source) provides a suitable calibration source for those objects.
However, there is no similar source that is bright, has a constant flux and with a known, extended morphology available to act as an extended calibration source.

We present a novel method of generating a fake extended source by conducting a raster scan over the Crab Nebula, this generates an extended source of known flux and morphology allowing for the verification of extended source analyses.
This method is complementary to existing techniques where a simulated source is added to a true background, since it is less reliant on a detailed understanding of the response of the systems away from the centre of the field of view.
This work presents initial proof of principle tests to show that the method works for the generation of extended sources and for testing the existing extended source analysis used by VERITAS.
In the future, by varying the scan (or applying time cuts to an existing scan) it will be possible to generate a wide variety of source morphologies allowing for custom verification of sources and analysis testing.
It will be particularly useful for objects that are expected to show significant morphology (e.g. supernova remnants) or those that present significant analysis challenges, for example Geminga which, as detected by Milagro \cite{Abdo2009a}, is expected to almost fill the VERITAS field of view.

\section{Methodology}
The raster scan is adapted from the software used for the VERITAS mirror alignment \cite{McCann2010} and the star raster scan for relative throughput measurements \cite{Griffin2013}.
It works by adjusting the pointing corrections of telescopes, thus the telescope still believes that it is pointing in the original direction though it now has an offset.
This allows for the standard analysis method \cite{Daniel2007} to be used with no changes, (note that the VERITAS pointing monitors - a set of optical star trackers that are used to correct for any miss-pointing the the telescopes - are not used else they will correct for the offset introduced by the scan).
The pattern followed by the scan can be adjusted and the code has been set up so the four telescopes follow the same pattern at the same time.

The observations are conducted in the standard way, with the Crab Nebula offset from the camera center by \ang{0.5} in one of the cardinal directions (N, S, E or W) (\textit{Wobble} mode) to enable background estimation from the rest of the camera.
Data collection is then started before the raster scan commences and the run is longer than the length of the scan, this additional time at the start and the end of the run is removed using time cuts.
The pointing of the array is recorded for each raster position and used to determine the actual pattern followed.  
From that the apparent location of the Crab Nebula is determined.

In this initial test, two patterns that are loosely rectangular in shape were used, with different orientations and offset directions.
In the future, the raster program will be updated to generate a box in RA/Dec that will allow for multiple scans to be overlaid.

\section{Scan Patterns}
\begin{figure}[h]
\centering
\begin{subfigure}[t]{0.49\textwidth}
\includegraphics[width=1\linewidth]{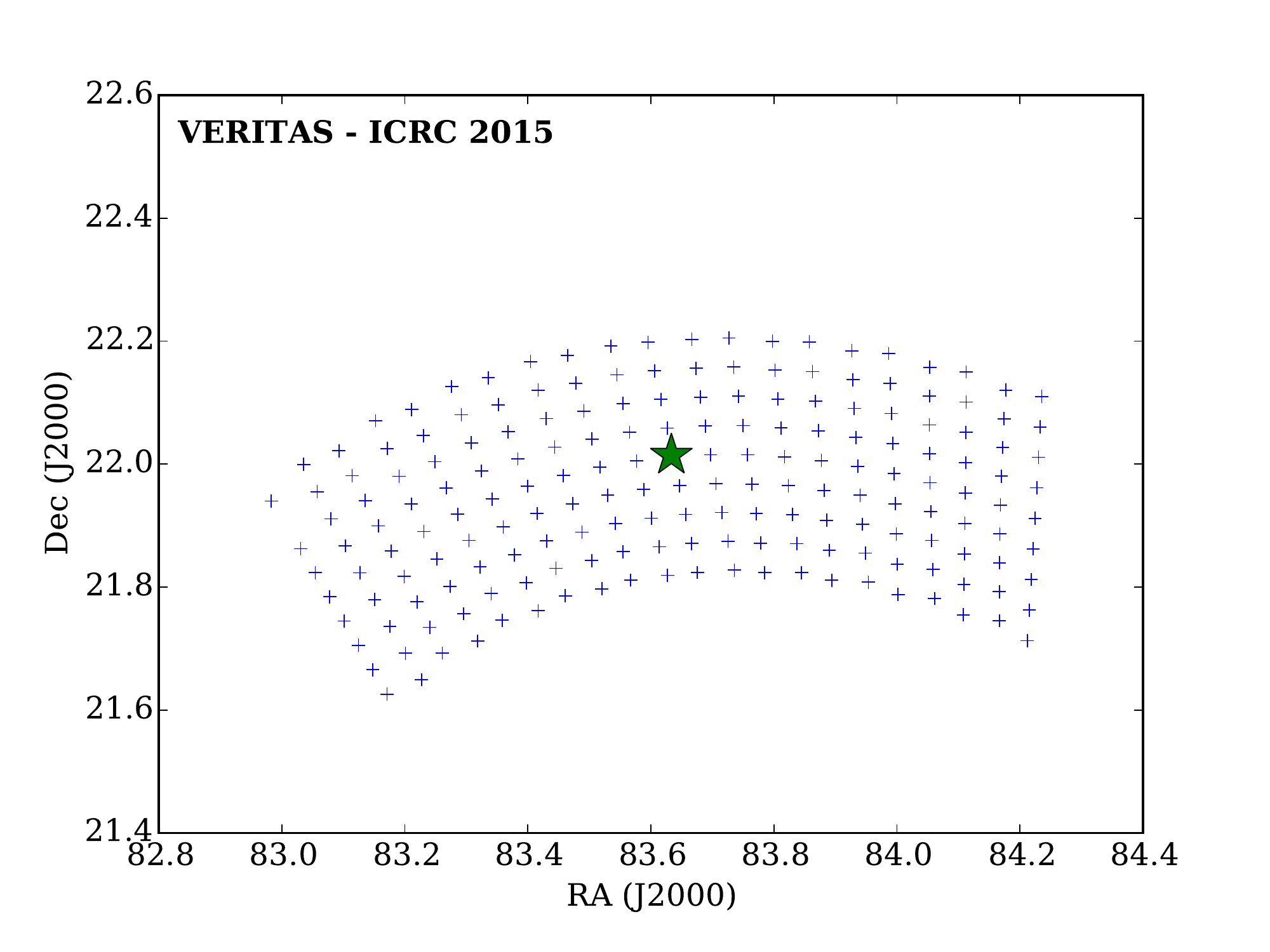}
\caption{\emph{Pattern 1}}
\label{Fig:rawScan1}
\end{subfigure}
\hfill
\begin{subfigure}[t]{0.49\textwidth}
\includegraphics[width=1\linewidth]{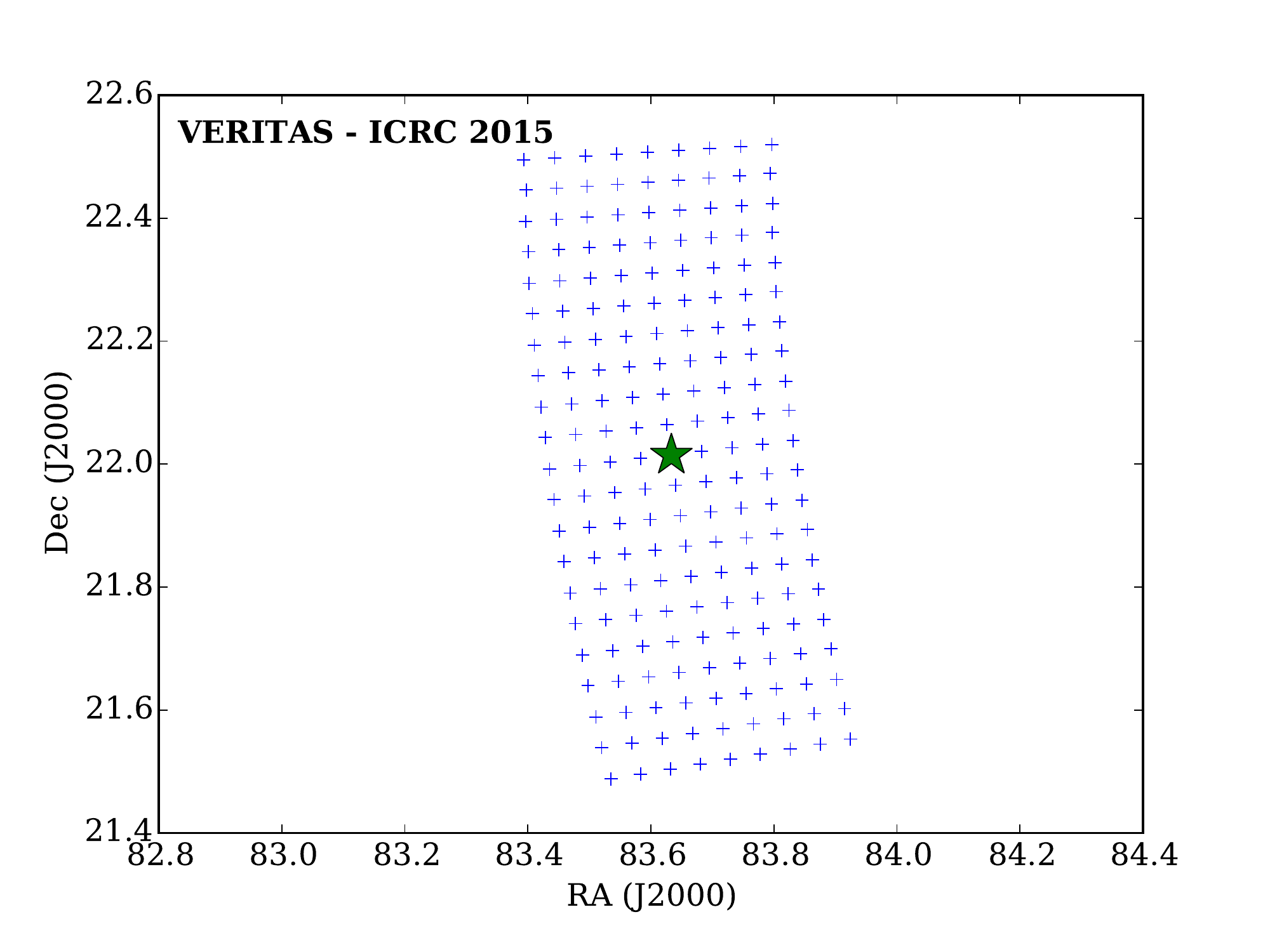}
\caption{Pattern 2}
\label{Fig:rawScan2}
\end{subfigure}
\caption{The raster patterns that were employed.  The large star indicates the Crab Nebula position, the blue crosses are the effective positions of the Crab Nebula for each point in the scan.}
\end{figure}

Two scan patterns were performed (Figures \ref{Fig:rawScan1}, one wobble N thus the scan does not cross the camera center, and \ref{Fig:rawScan2} and one wobble S where it does cross the camera center) with a dwell time of \SI{10}{s} on each point.  
With 189 pointings in total this gives a scan time of 31.5~minutes.
The pattern was designed with sufficient statistics and a noticeable asymmetry in the shape for verification that the scan has produced the desired result. 

With a ``standard'' point source analysis, using the same set up as used in this analysis (but without the extended exclusion regions) the Crab Nebula is detected at a significance of \similar40$\sigma$ per sqrt(hour), thus for a 30 minute run we would expect a detection at \similar30$\sigma$.
However, since the Crab Nebula will be ``smeared'' over an area, approx \ang{1} by \ang{0.5}, the fraction of the flux that falls within a test region will be significantly below this, thus though we expect to get a significant detection, it will be significantly lower than that from the point source.
An estimate of the expected significance can be made for the test region of radius \ang{0.3} centred on the true location of the Crab Nebula.
Using values determined from observations of the Crab Nebula conducted using the same analysis set up we expect a significance of around 12$\sigma$)

\section{Skymaps}
The standard VERITAS analysis is used with standard gamma/hadron selection cuts to produce skymaps using the ring background method (where the background region is taken from a ring surrounding the test position, \cite{Berge2007}).
In generating the skymap the whole of the area covered by the effective Crab Nebula positions and identical area centred on $\zeta$-Tauri are excluded from the background model.
For the weaker stars in the field-of-view only the nominal positions are excluded since a similar large exclusion region significantly reduces the background available and the impact of the star is diminished since its effect is smeared out over a larger area.

\begin{figure}[h]
\centering
\begin{subfigure}[t]{1.\textwidth}
\centering
\includegraphics[trim=15mm 0mm 15mm 0mm, clip=true, width=0.75\linewidth]{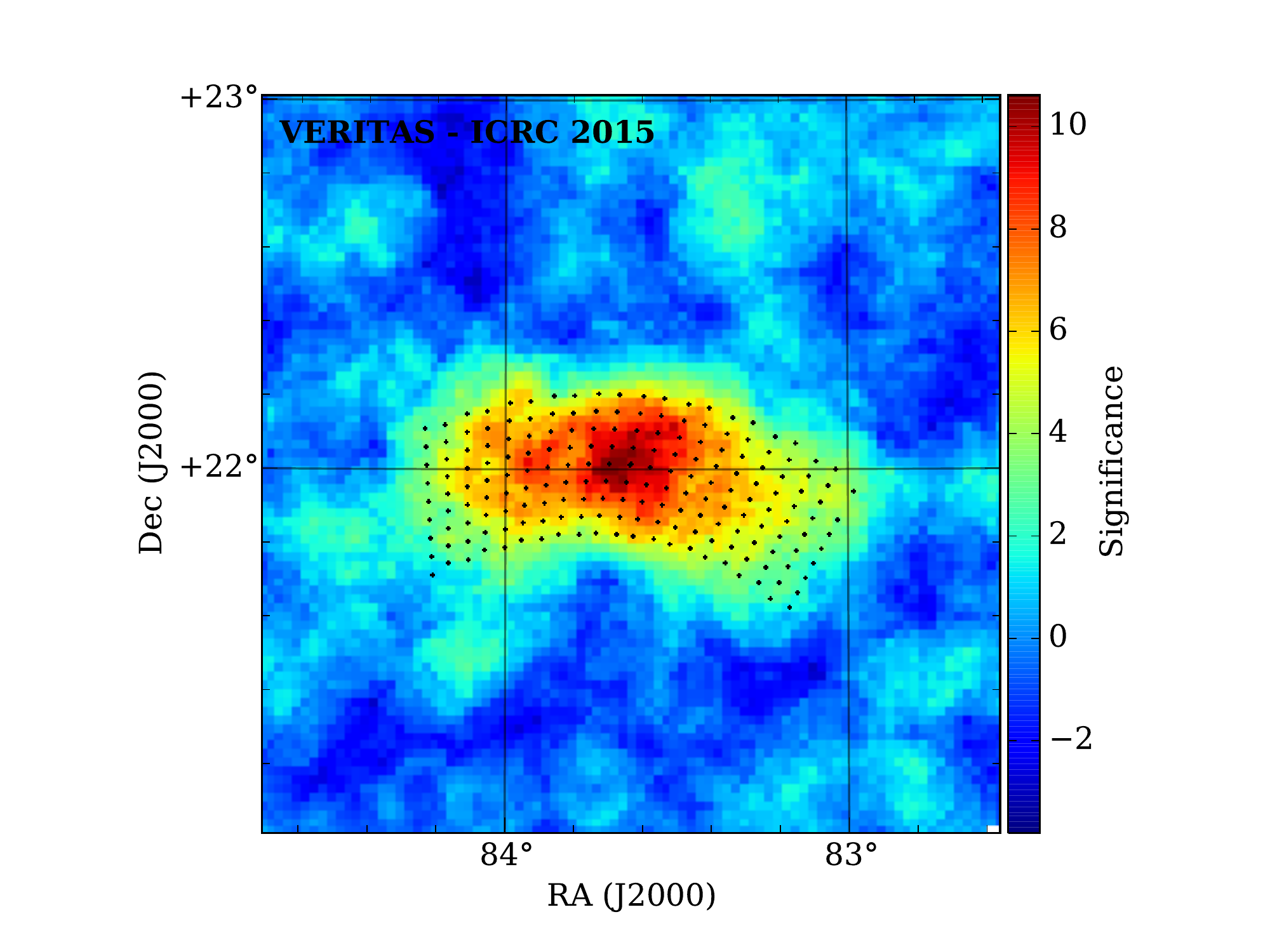}
\caption{Test region radius = \ang{0.1}}
%\label{Fig:skymap1}
\end{subfigure}
\begin{subfigure}[t]{1.\textwidth}
\centering
\includegraphics[trim=15mm 0mm 15mm 0mm, clip=true, width=0.75\linewidth]{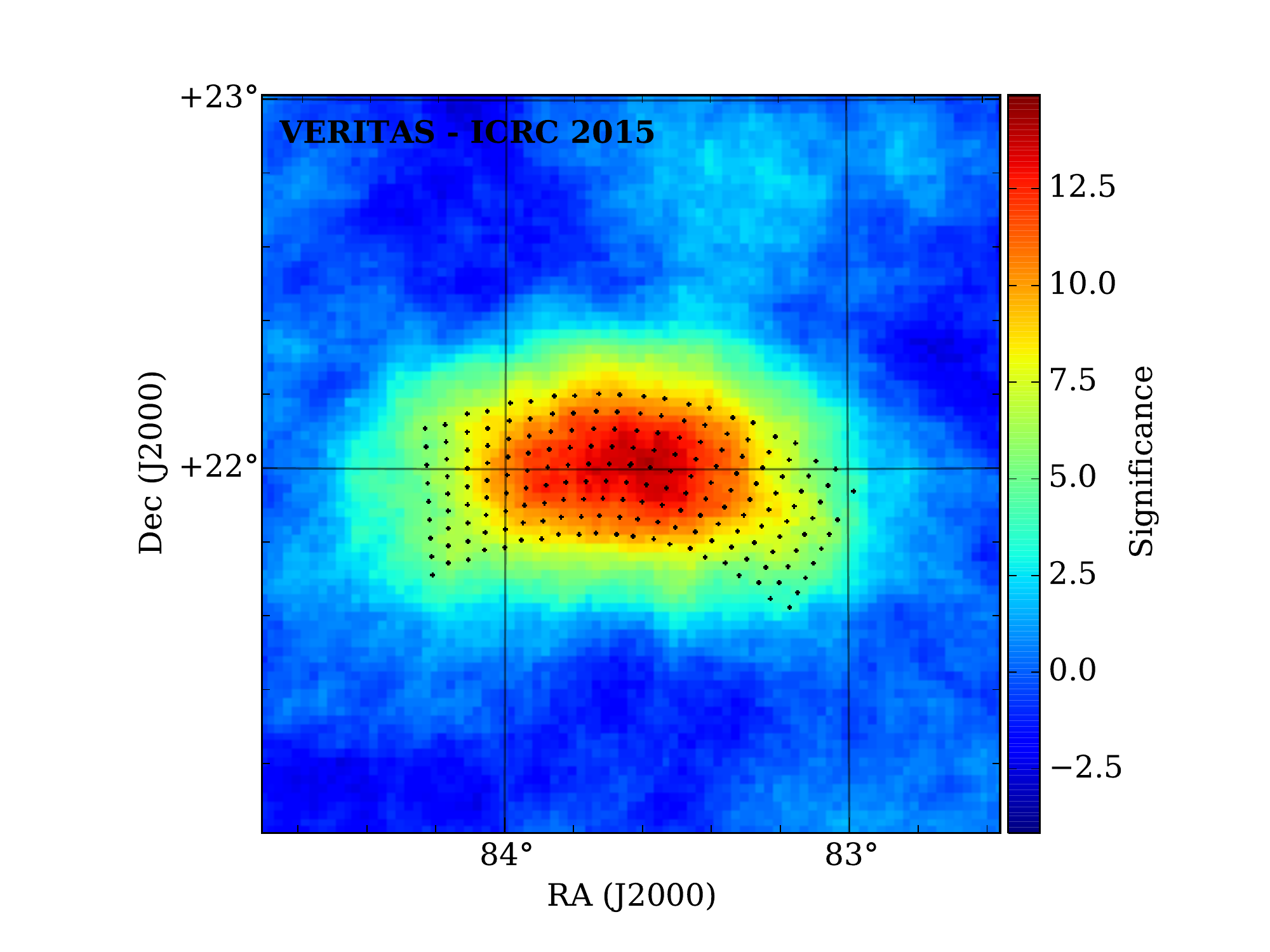}
\caption{Test region radius = \ang{0.3}}
%\label{Fig:skymap1VExt}
\end{subfigure}
\caption{Skymap of the Crab Raster Scan for \emph{Pattern 1}, the nominal Crab positions are the black dots.}
\label{Fig:PointSky}
\end{figure}

\begin{figure}[h]
\centering
\begin{subfigure}[t]{1.\textwidth}
\centering
\includegraphics[trim=15mm 0mm 15mm 0mm, clip=true, width=0.75\linewidth]{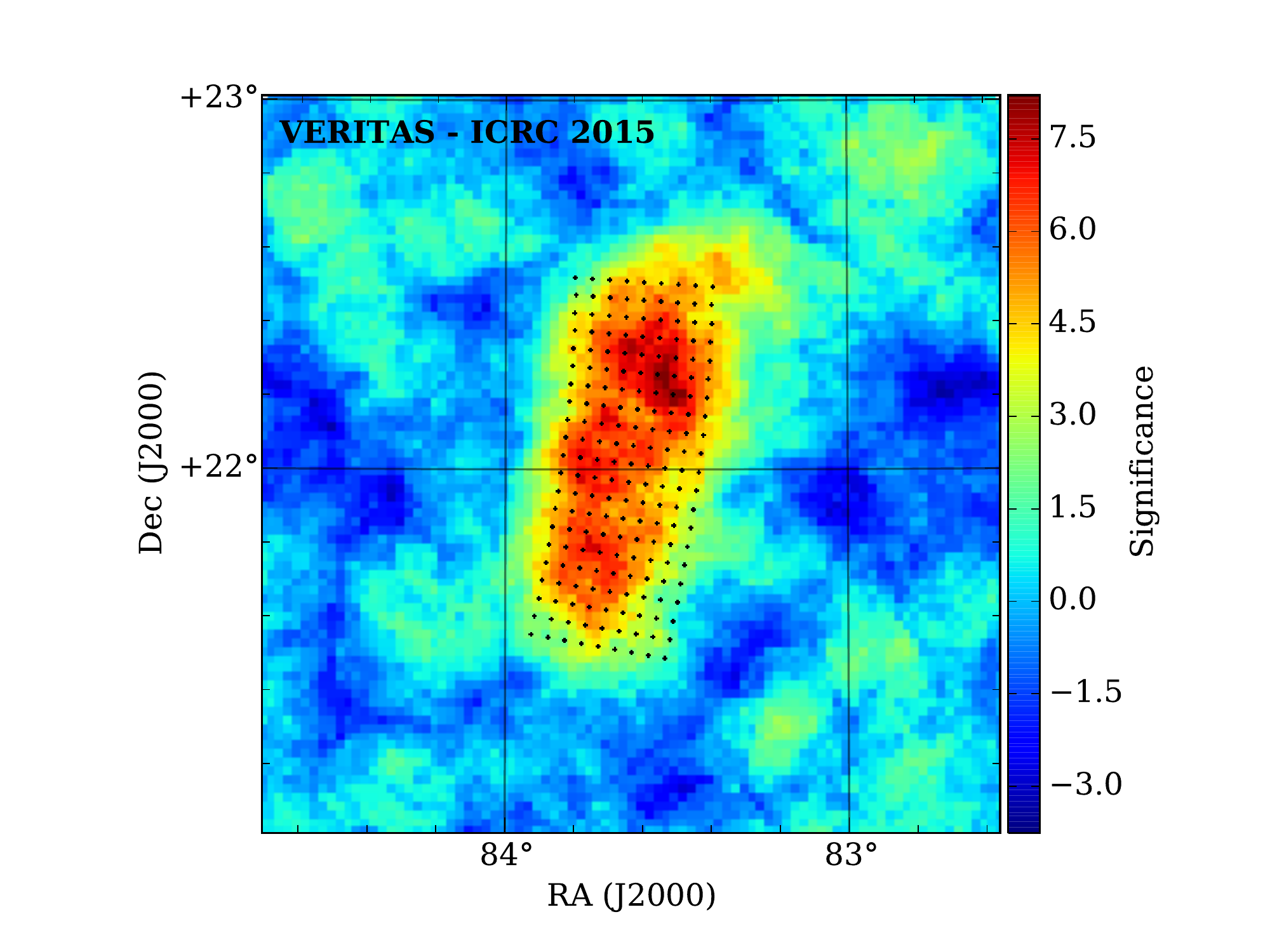}
\caption{Test region radius = \ang{0.1}}
%\label{Fig:skymap2}
\end{subfigure}
\begin{subfigure}[t]{1.\textwidth}
\centering
\includegraphics[trim=15mm 0mm 15mm 0mm, clip=true, width=0.75\linewidth]{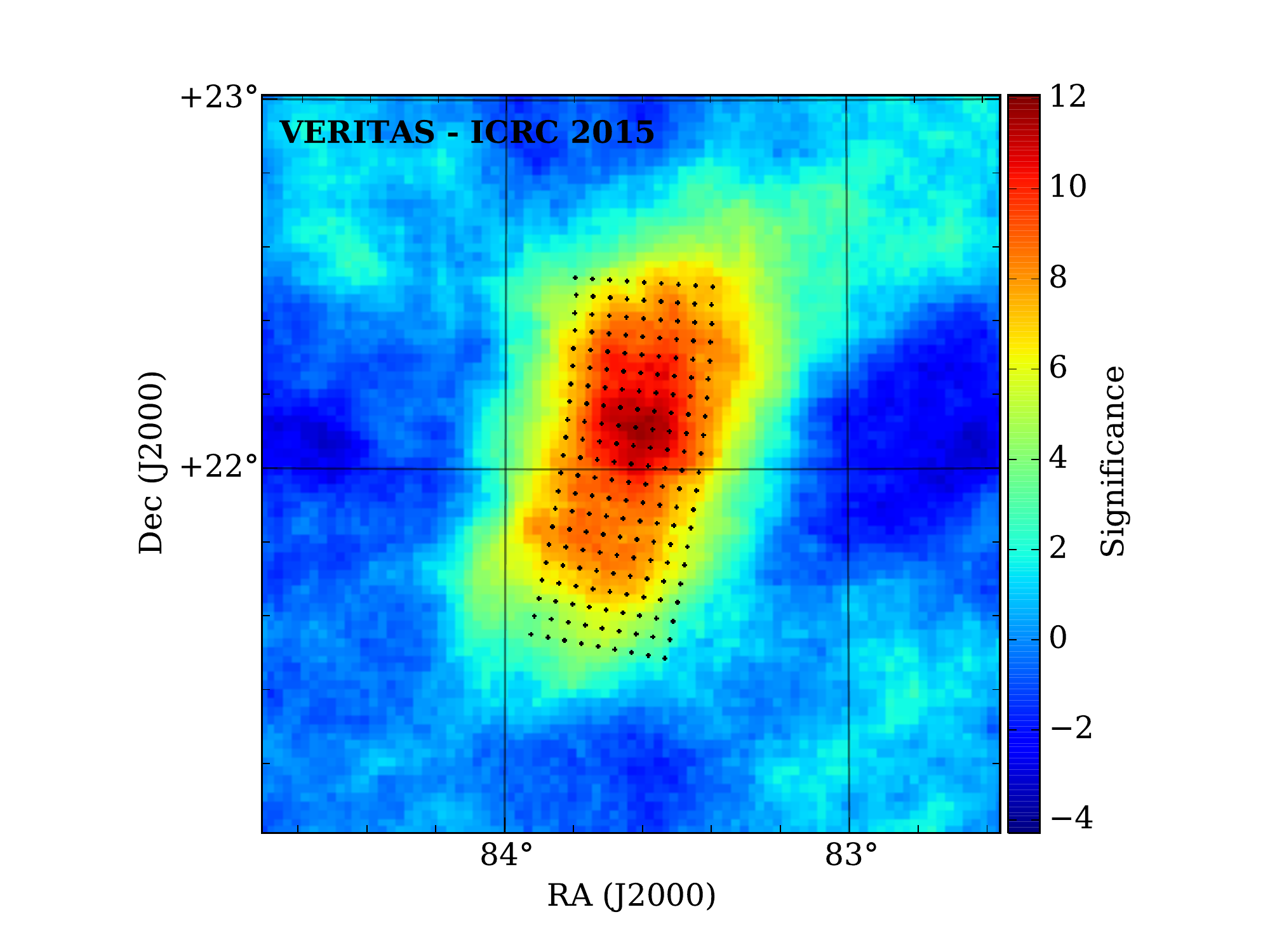}
\caption{Test region radius = \ang{0.3}}
%\label{Fig:skymap1VExt}
\end{subfigure}
\caption{Skymap of the Crab Raster Scan for \emph{Pattern 2}, the nominal Crab positions are the black dots.}
\label{Fig:VExtSky}
\end{figure}

Two different test region radii are tested, \ang{0.1} and \ang{0.3}, the skymaps of these for each of the patterns are shown in Figures \ref{Fig:PointSky} and \ref{Fig:VExtSky}.
They clearly show an extended source that is coincident with the shape scanned out by the raster process, thus showing that the method has effectively produced an extended test source.
The peak significance in the maps with a test region radius of \ang{0.3} is 10-12$\sigma$, a reasonable agreement with the simple calculation above, though there are noticeable variations across the scan due to statistical fluctuations.

The edges of the scan, where the statistics are lower, show larger statistical fluctuations, this is ``smeared out'' by the larger test regions creating a smoother shape but they have also had the effect of reducing the significance of the edges of the scan.
This reduces the apparent size of the source with the high significance region significantly away from the ends of the scan.
The significance of the detection also depends upon the size of the test region, with a larger test region containing a larger signal, increasing the significance of the detection.
If the test region is made larger still then the significance decreases due to the reduction in the background area available (with a test region radius of \ang{0.5} the peak significance for \textit{Pattern 2} drops to about 8).

\section{Spectral Reconstruction}
\begin{figure}[h]
\centering
\includegraphics[trim=0mm 0mm 15mm 0mm, clip=true, width=0.95\linewidth]{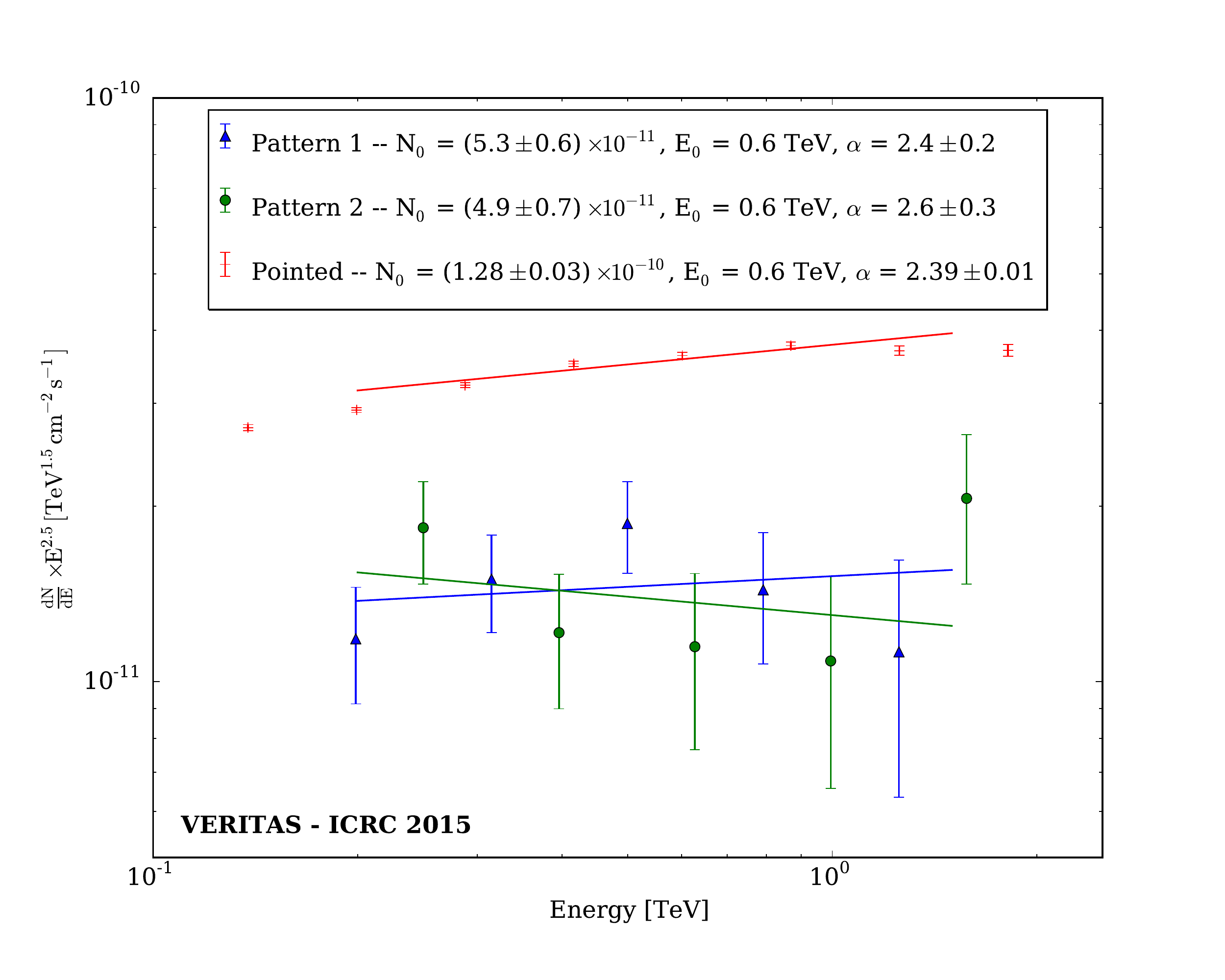}
\caption{Spectrum from the center of the scan in \emph{Pattern 1} - red and \emph{Pattern 2} - blue compared with the VERITAS Crab Nebula spectrum fitted over a the energy range $0.2 - \SI{1.5}{TeV}$, \emph{Pointed} - black.  This analysis was conducted with a test region radius of \ang{0.3}}
\label{Fig:VExtSpec}
\end{figure}

\begin{sloppypar}
A spectral analysis of the region centred on the true Crab position (the central point of the scan) is shown in Figure \ref{Fig:VExtSpec}.
Each spectrum was fit with a power law of the form $\frac{\diff N} {\diff E} = N_0 \left(\frac{E}{\SI{0.6}{TeV}}\right)^{-\alpha} \si{TeV^{-1} cm^{-2} s^{-1}}$.
The \textit{Pointed} spectrum is the taken from the VERITAS spectrum presented in \cite{Kevin} but with the fit conducted over the range 0.2 to \SI{1.5}{TeV}.
\end{sloppypar}

\begin{table}[h]
\centering

\begin{tabular}{cccc}
\hline
Run & Significance [$\sigma$] &$N_0$ & $\alpha$ \\
\hline
\emph{Pointed}   & 15.62 & \num{1.28 \pm 0.04e-10} & \num{2.30 \pm 0.01}  \\
\emph{Pattern 1} & 11.68 & \num{5.3 \pm 0.6e-11} & \num{2.4 \pm 0.2} \\
\emph{Pattern 2} & 8.93 & \num{4.9 \pm 0.7e-11} & \num{2.6 \pm 0.3}  \\
\hline
\end{tabular}
\caption[Crab Nebula UVF + RHV Observations]{Crab nebula UVF + RHV observations \similar~\ang{3} from a 45\% illuminated Moon.}
\label{Tab:UVFRHVCrab}
\end{table}

As expected the fluxes from the raster observations are significantly lower than for the standard Crab Nebula observation since only a fraction of the flux now ``originates'' within the test region.

An estimate of the fraction of the flux that we would expect can be made by convolving the raster scan patter with the VERITAS point spread function (\ang{0.1}) and determining the amount that falls within the test region out of the total flux from the whole object.
Correction also has to be made for the fact that the effective area is generated using a point source whereas this observation is of an extended source.
These calculations show that we would expect the measured flux to be (\emph{Pattern 1}/\emph{Pattern 2}) 46.8\%/50.8\% of the total flux from the entire source.
Applying these scaling factor to the measured normalisation gives an estimate of $N_0$ from the entire scan of \num{1.1 \pm 0.1 e-10}/\num{9.8 \pm 1.4 e-11}, these are close to the nominal value.

In the future, by taking more scans and stacking the results the statistical errors will reduce significantly, this will enable more detailed testing of this method of estimating the flux from the entire object and further verification of the extended source analysis.

\section{Conclusions}
\begin{sloppypar}
By conducting a raster scan across the Crab Nebula it is possible to generate an extended source.
Since the scan follows a known pattern the morphology of the source is known and thus it can be used for testing extended source analysis methods.
The central, \ang{0.3} radius region within each scan was fit with a power law of $N_0$ (\emph{Pattern 1}/\emph{Pattern 2}) = \num{5.3 \pm 0.6e-11}/\SI{4.9 \pm 0.7e-11}{TeV^{-1} cm^{-2} s^{-1}}, $\alpha$  = \num{2.4 \pm 0.2}/\num{2.6 \pm 0.2}, E$_0$ = \SI{0.6}{TeV}.
Comparison with a model of the scan showed that this was expected to be 46.8\%/50.8\% of the total flux from the scan, thus the $N_0$ for the whole region is estimated at \num{1.1 \pm 0.1 e-10}/\SI{9.8 \pm 1.4 e-11}{TeV^{-1} cm^{-2} s^{-1}}.
For comparison, the VERITAS Crab Nebula spectrum fitted over the same energy range has a normalisation of \SI{1.28 \pm 0.04e-10}{TeV^{-1} cm^{-2} s^{-1}} and an index of \num{-2.30 \pm 0.01}.
With improvements to the raster code, multiple scans will be combined to improve the statistics of the method and enable its use for improved testing of spectral reconstruction and morphology for extended objects.
\end{sloppypar}

\acknowledgments
This research is supported by grants from the U.S. Department of Energy Office of Science, the U.S. National Science Foundation and the Smithsonian Institution, and by NSERC in Canada. 
We acknowledge the excellent work of the technical support staff at the Fred Lawrence Whipple Observatory and at the collaborating institutions in the construction and operation of the instrument.
R. Bird is funded by the DGPP which is funded under the Programme for Research in Third-Level Institutions and co-funded under the European Regional Development Fund (ERDF).
The VERITAS Collaboration is grateful to Trevor Weekes for his seminal contributions and leadership in the field of VHE gamma-ray astrophysics, which made this study possible.


\begin{thebibliography}{99}
\bibitem{OFDB2015}
O'Faol\'ain de Bhr\'oithe, A., et al., (2015). \emph{The search for short-term flares in extended VHE Crab Nebula observations with the Whipple 10 m telescope}. these proceedings

\bibitem{Abdo2009a}
Abdo, A. A., et al., (2009). \emph{MILAGRO observations of multi-TeV emission from Galactic sources win the Fermi Bright Source List}. ApJ, 700(2), L127-L131. 

\bibitem{McCann2010}
McCann, A., et al., (2010). \emph{A new mirror alignment system for the VERITAS telescopes}. Astroparticle Physics, 32(6), 325-329. 

\bibitem{Griffin2013}
Griffin, S., \& Hanna, D. (2013). \emph{Using Raster Scans of Bright Stars to Measure the Relative Total Throughputs of Cherenkov Telescopes. Instrumentation and Methods for Astrophysics}. in \emph{Proceedings of the 33rd ICRC}

\bibitem{Daniel2007}
Daniel, M. K., et al., (2007). \emph{The VERITAS standard data analysis}. in \emph{Proceedings of the 30th ICRC}

%\bibitem{Li1983}
%Li, T., \& Ma, Y. (1983). \emph{Analysis methods for results in gamma-ray astronomy}. ApJ, 272, 317.

\bibitem{Berge2007}
Berge, D., Funk, S., \& Hinton, J. (2007). \emph{Background modelling in very-high-energy \GRs astronomy}. A\&A, 466(3), 1219-1229. 

\bibitem{Kevin}
Meagher, K., et al., (2015). \emph{Six years of VERITAS observations of the Crab
Nebula}. these proceedings
\end{thebibliography}
\end{document}